\documentclass{ws-jai}
\usepackage[flushleft]{threeparttable}
\usepackage[colorinlistoftodos]{todonotes}
\usepackage{hyperref}

\def\sk{$\widehat{SK}$~}
\def\esk{\widehat{SK}}
\def\pasp{Publications of the Astronomical Society of Pacific}

\def\apjl{The Astrophysical Journal Letters}
\def\mnras{Monthly Notices of the Royal Astronomical Society}

\def\nat{Nature}

\begin{document}
\catchline{}{}{}{}{} 
\markboth{Nita et. al}{Statistics of 2-bit digitized signals}
\title{Statistical discrimination of RFI and astronomical transients in 2-bit digitized time domain signals}

\author{Gelu M. Nita$^1$, Aard Keimpema$^2$, Zsolt Paragi$^2$}

\address{
$^1$Center for Solar-Terrestrial Research, New Jersey Institute of Technology, Newark, NJ 07102, USA, gnita@njit.edu\\
$^2$Joint institute for VLBI ERIC}

\maketitle
\begin{abstract}
We investigate the performance of the generalized Spectral Kurtosis (SK) estimator in detecting and discriminating natural and artificial very short duration transients in the 2-bit sampling time domain Very-Long-Baseline Interferometry (VLBI) data. We demonstrate that, while both types of transients may be efficiently detected, their natural or artificial nature cannot be distinguished if only a time domain SK analysis is performed. However, these two types of transients become distinguishable from each other in the spectral domain, after a 32-bit FFT operation is performed on the 2-bit time domain voltages. We discuss the implication of these findings on the ability of the Spectral Kurtosis estimator to automatically detect bright astronomical transient signals of interests -- such as pulsar or fast radio bursts (FRB) -- in VLBI data streams that have been severely contaminated by unwanted radio frequency interference.

\end{abstract}








\section{Introduction}
\label{sec:intro}

Radio frequency interference (RFI)  mitigation algorithms based on statistical discrimination of signals in the frequency or time domains \citep{Fridman2001, FB2001, Ruf} have become increasingly popular in recent years, due to the feasibility of their implementation using field-programmable gate arrays (FPGAs) or graphical processing units (GPUs), equipped on many modern digital instruments \citep{Baan2004,sks}.

Amongst these methods, the Spectral Kurtosis estimator \citep[$\widehat{SK}$,][]{rfi} has been proven to be very effective in detecting RFI contamination of astronomical signals, regardless their temporal or spectral profiles. This is because, a remarkable property of the \sk estimator is that, in the case of a pure Gaussian time domain signal, its statistical expectation is unity at each frequency bin, while the power spectrum may have an arbitrary shape. This property gives to the \sk estimator the ability to discriminate signals deviating from a Gaussian time domain statistics against arbitrarily shaped natural astronomical signals obeying it, as it is usually the case of the man-made signals producing unwanted RFI contamination of astronomical signals of interest. This ability confers to the \sk estimator a net advantage over alternative RFI excision algorithms that solely employ power-based empirical thresholding algorithms \citep[e.g. AOFlagger,][]{Offringa2012},  which ignore the higher order statistical information encoded in the signals, and thus are prone to mistakenly interpret any detected astronomical transients as RFI contamination \citep{Cendes2018}.

A simple design for a wide-band SK spectrometer was first implemented and tested in a software correlator \citep{FST} and, subsequently, in the hardware correlators of two prototype instruments \citep{jgr}. Since then, the SK spectrometer design has been embedded in the hardware correlators of two fully operational instruments, namely the Korean Solar Burst Locator \citep[KSRBL,][]{ksrbl, sks}, and the Expanded Owens Valley Solar Array  \citep[EOVSA,][]{jai}.

\citet{sk} updated the expression of the \sk estimator proposed in \citet{rfi} to remove the statistical bias of the original definition and, furthermore, \citet{gsk} extended its applicability to any signal stream $x_i$ obeying a gamma probability distribution characterized by a shape parameter $d$ and a scale parameter $\lambda$, denoted from now on as $\Gamma(d,\lambda)$.

The generalized \sk estimator introduced by \citet{gsk},
\begin{equation}
\label{gsk}
\esk=\frac{M d+1}{M-1}\Big(\frac{M S_2}{S_1^2}-1\Big),
\end{equation}
were $S_1=\Sigma_{i=1}^M x_i$ and $S_2=\Sigma_{i=1}^M x_i^2$, describes the statistical properties of a series of power spectral density (PSD) estimates $x_i$ obtained by means of narrow-band time-domain filtering ($d=0.5$, chi-square distribution with one degree of freedom), or Fast Fourier Transform (FFT) ($d=1$, exponential distribution) \citep{rfi, sk, gsk}, as well as the statistical properties of a set of already integrated PSD estimates over an accumulation length $N$,  since such sums of $\Gamma(d,\lambda)$ distributed random variables are expected to obey a $\Gamma(N d,\lambda)$ statistics. Therefore, the generalized \sk estimator may be employed, in both time and frequency domains, not only with data produced by a new generation of instruments specifically designed for its use, but also by any already existing classical instruments that, by design, only output already integrated PSD estimates \citep{gsk, jai}.

Moreover, as demonstrated by \citet{qsk} using filter bank data from the CSIRO Parkes telescope \citep{Parkes}, the generalized \sk estimator given by Equation~\ref{gsk} may be also used to describe the statistical properties of a set of already integrated PSD estimates recorded in a 2-bit quantized format, a case in which the statistical distribution of such quantized PSD estimates may be satisfactorily approximated by a gamma distribution characterized by an instrument-specific shape factor that can be empirically determined from the data.

Here we complement and detail a preliminary study that we briefly reported in \citet{globalsip}, by presenting a comprehensive time and spectral domain data analysis, supported by numerical simulations, that addresses the applicability and performance of the generalized \sk estimator in the case of 2-bit quantized voltages recorded by instruments in the European VLBI Network \citep[EVN,][]{2015arXiv150105079Z}, not only as a RFI detector, but also, more generally, as a statistical tool that, as theoretically shown by \citet{Nita2016} and experimentally tested by \citet{jgr}, may be used to automatically detect and infer the natural or artificial nature of radio transients, an aspect of general astronomy interest in the context of studies searching for pulsars, fast radio bursts \citep[FRB,][]{Lorimer2007}, or extra terrestrial intelligence (SETI).

To do so, in \S\ref{sec:instr} we present an overview of the VLBI instrument and data, in \S\ref{sec:VLBI_TDK} we perform a time domain statistical analysis of data segment capturing the FRB 121102 \citep{Spitler2016}, in \S\ref{sec:sim} we support the interpretation of our findings through numerical simulations, in \S\ref{sec:VLBI_SK} we extend our analysis of the same data segment to the spectral domain by replicating and detailing the spectral domain analysis presented in \citet{globalsip}, and we conclude our study in \S\ref{sec:conclusions}.

\section{Overview of the VLBI Instruments and Data}
\label{sec:instr}

Real-time-processing (correlation) electronic-VLBI operations made the e-EVN array more flexible to follow-up transient phenomena \citep{Paragi2016}. The data presented here have been streamed from the 305-m William E. Gordon Telescope at the Arecibo Observatory and recorded at the Joint Institute for VLBI ERIC (JIVE), during the e-EVN monitoring campaign in the active period of FRB\,121102 on 20 September 2016. This is so far the only FRB that has produced multiple bursts \citep{Spitler2016}. The e-EVN telescopes participating in the campaign were Arecibo, Effelsberg, Jodrell Bank (MkII), Westerbork (single dish), Medicina and Onsala (25m). The resulting data allowed the first-ever VLBI detection and the precise measurement of the position of a fast radio burst \citep{Marcote2017}, following the initial localization of FRB\,121102 with the JVLA \citep{Chatterjee2017}.

The central observing frequency was 1.7 GHz. Most telescopes streamed 2-bit sampled data at a data rate of 1024 Mbit/s, while from Arecibo the network limited us to stream 512 Mbit/s. The latter corresponds to 64~MHz bandwidth in two polarizations (left- and right-hand circular), that were divided into four 16~MHz sub-bands. The required sampling rate is 32 Msamples/s/sub-band, resulting in 31.25~ns time samples.

The pulses of FRB\,121102 are dispersed as they travel through the ionized interstellar and intergalactic medium. The dispersion measure is DM=557 parsec cm$^{-3}$. This results in a frequency-dependent delay of pulse arrival time between the sub-bands, and a time smearing of about 20~ms within each sub-band, as seen in Figures \ref{fig:LCP_TDK} and \ref{fig:RCP_TDK}, unless the data are dedispersed. In total, four pulses were detected in the observation, each pulse having an intrinsic duration, i.e. after dedispersion, of $\sim$2~ms.

\section{Time Domain Spectral Kurtosis Analysis of FRB 121102}
\label{sec:VLBI_TDK}

To perform a generalized \sk analysis of the time domain FRB 121102 data, we chose to produce uncalibrated power and power squared estimates by simply computing the seconds and fourth powers of the raw voltages for each of the four wide-band spectral sub-bands and both circular polarizations, which were subsequently integrated over contiguous time blocks of M=11520 data samples (0.36~ms time bins) to form the sums $S_1$ and $S_2$ entering the expression of the \sk estimator given by Equation \ref{gsk}.

\begin{figure*}[htb]
\begin{minipage}[b]{1.0\linewidth}
  \centering
  \centerline{\includegraphics[width=17.5cm]{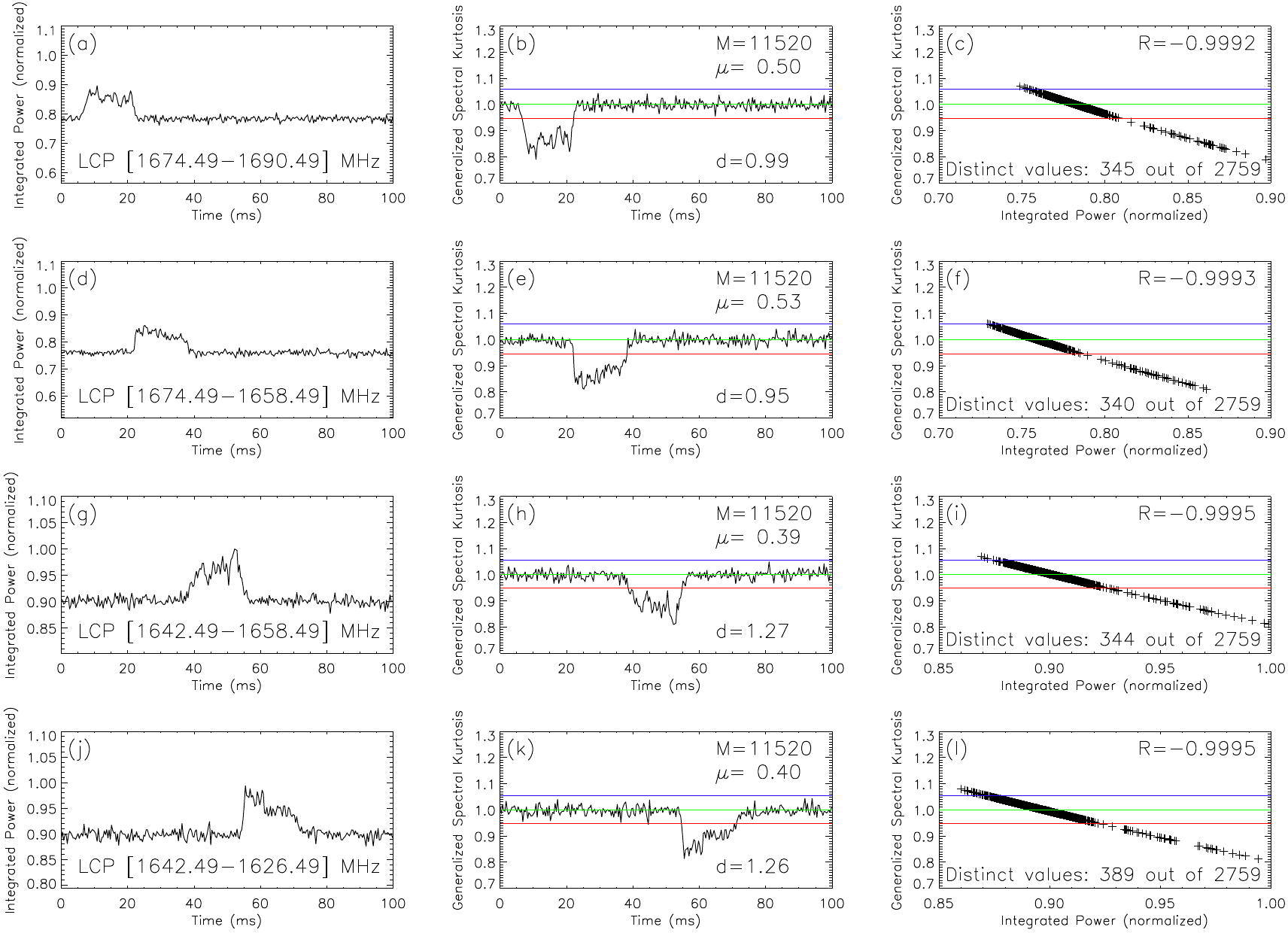}}
 \caption{Time domain analysis of FRB 121102 signal observed by VLBI in four $\sim$16MHz-width spectral sub-bands with a 31.25~ns sampling rate. Left column: Normalized time series of LCP power estimates integrated over M=11520 2-bit sampling contiguous voltage  (0.36~ms time bins). Middle column: Generalized Spectral Kurtosis computed according to Equation~\ref{gsk} assuming a sub-band-dependent shape factor $d$ estimated from the data, indicated in each panel inset. The mean values $\mu$ used to normalize the $d=0.5$ \sk estimators to unity are shown in each panel inset. Right column: Spectral correlation vs. Integrated power. An almost perfect negative correlation revealed, The correlation coefficients $R$ corresponding to each spectral sub-band are indicated in each panel inset. The number of distinct SK realizations corresponding to a total 2759 accumulations are displayed.}
\label{fig:LCP_TDK}
\end{minipage}
\end{figure*}

\begin{figure*}[htb]
\begin{minipage}[b]{1.0\linewidth}
  \centering
  \centerline{\includegraphics[width=17.5cm]{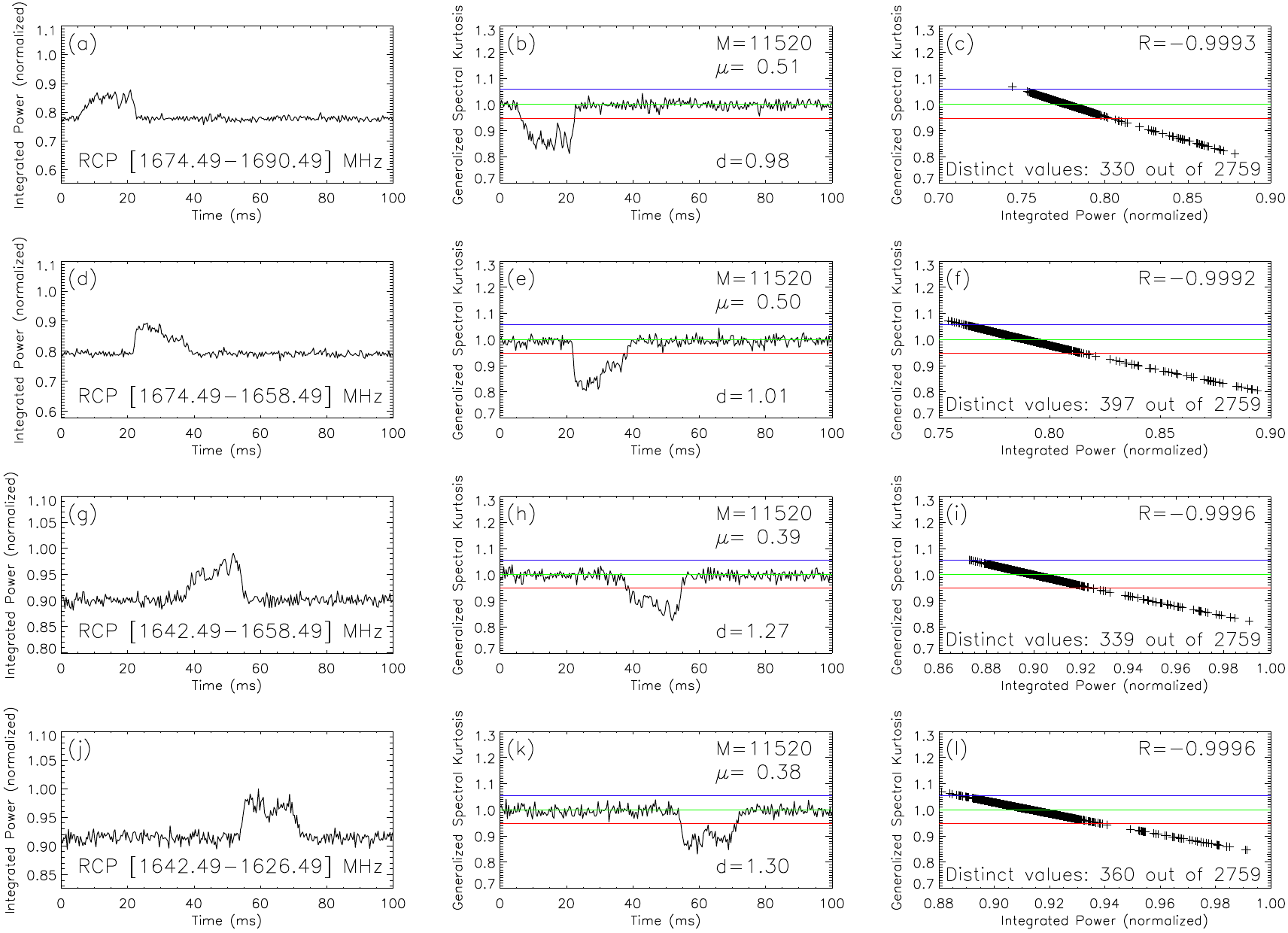}}
 \caption{Time domain analysis of FRB 121102 VLBI 2-bit sampling RCP voltage data. The panel organization is the same as in Figure~\ref{fig:LCP_TDK}.}
\label{fig:RCP_TDK}
\end{minipage}
\end{figure*}

The left-hand columns in Figures~\ref{fig:LCP_TDK} and~\ref{fig:RCP_TDK} display the accumulated power estimates $S_1$ corresponding to the left-hand (LCP) and right-hand (RCP) circular polarizations, respectively, over a time segment of 100~ms originating right before the onset of the FRB signal in the [1674.49-1690.49]~MHz raw voltage sub-band. The drifting FRB 121102 signal is evident in all these time series and, since its particularly high signal to noise ratio (SNR) and flat background does not pose any challenge to its detection through simple $S_1$ empirical thresholding, we concentrate here just on the statistical properties of the recorded signals, as resulted from \sk time series shown in the middle-panels of both Figures~\ref{fig:LCP_TDK} and~\ref{fig:RCP_TDK}.

To compute the \sk estimators shown in these panels, we first assume that the squared raw voltage samples, in the time regions outside the active FRB signal, obey a ChiSqr distribution, which is the theoretical expectation for a Gaussian time domain signal recorded with (practically) infinite bit sampling. Thus, employing Equation \ref{gsk} with M=11520 and $d=0.5$ (ChiSqr distribution), we obtained \sk time series having, outside the time range containing the active FRB signal, the sub-band dependent non-unity sample means $\mu$ shown in each panel inset, i.e. $\mu=\{0.50,0.53,0.39,0.40\}$ and $\mu=\{0.51,0.50,0.39,0.38\}$ for the LCP and, respectively, RCP sub-bands.

The conclusion that can be inferred from these far from unity \sk means is that, although, in the absence of any transients or RFI contamination, the sub-band astronomical background signals are expected to obey a Gaussian distribution, their 2-bit quantized representation do not. However, the fact that the \sk values associated with the FRB transient signal show systematic deviations from the mean level of the background \sk values demonstrates that, despite the signal quantization, the \sk estimator defined by Equation \ref{gsk} preserves its ability to discriminate signals that do not obey the same statistics as the background level.

Therefore, assuming that the true statistical distributions obeyed by the VLBI squared voltages may be approximated by gamma distributions, we use the sample means of the $d=0.5$ \sk estimators to infer from Equation \ref{gsk} the subband-dependent gamma distribution shape factors $d$ needed to normalize these \sk estimators to unity. This assumptions allows us to directly apply the theory of the generalized \sk estimator \citep{gsk} to compute asymmetric, but equal 0.13\% probability of false alarm (PFA) non-Gaussianity thresholds above (blue lines) and below (red lines) the unity \sk mean (green lines).

As illustrated in the in middle panels of both Figures~\ref{fig:LCP_TDK} and \ref{fig:RCP_TDK}, this procedure results in successful detection of the bright FRB\,121102 pulse, which, in all sub-bands, crosses the less than unity detection threshold.

We make, however, the cautionary note that, given the fact that the 2-bit quantization of the raw voltages results only in two possible distinct values of their squared values (1-bit quantization), which is an even more restrictive case than the 2-bit spectral domain quantification case analyzed by \citet{qsk}, the theory of the generalized \sk estimator developed by \citet{gsk} under the assumption of a continuous distribution of observable would need to be reconsidered.

Therefore, although the exact statistical properties of an \sk estimator based on flip-coin like discreet observables would need to be rigorously inferred from discrete statistics considerations involving the binomial distribution, which we do not attempt here, to illustrate an important effect that the quantization has on the behavior of the \sk estimator, we display in the right panels of Figures~\ref{fig:LCP_TDK} and \ref{fig:RCP_TDK} the relationship between the relatively small number of distinct \sk and $S_1$ realizations, which reveals an almost perfect negative correlation of these two magnitudes, regardless their location within or outside the the range delimited by the upper and lower non-Gaussianity thresholds. This \sk--$S_1$ correlation contrasts with a key property of the continuous \sk estimator, which is the statistical independence of the \sk and $S_1$ random variables associated with a time domain Gaussian signal \citep{sk,gsk}.

Therefore, our VLBI time series analysis indicates that the quantization process may limit the performance of the quantized \sk estimator as an automatic, unattended non-Gaussian signal detector, since its normalization to unity relies on the assumption of a constant mean background level, as opposed to its continuous counterpart, in which case its normalization is preserved as long as the variance of Gaussian background only changes at time scales longer than the integration time \citep{rfi}. Nevertheless, automatic gain control or automatic renormalization of the quantized \sk estimator could be in principle implemented in a time-domain data pipeline designed to automatically detect RFI contamination of  a quasi-stationary astronomical signal, as well as astronomical transient signals, such as pulsars, FRB, or even extra-terrestrial artificial transients.

However, neither of these two possible approaches to preserve its normalization would be able to overcome another important limitation revealed by the analysis of this particular VLBI data segment, which affects the ability of such time-domain quantized \sk estimator to automatically discriminate the artificial or natural origin of the detected transient signals based on the direction in which \sk deviates relative to the unity expectation corresponding to a time domain quasi-stationary Gaussian background.

Indeed, as demonstrated in a detailed study focused on the statistical properties of the transient signals \citep{Nita2016}, the \sk estimator associated with a normally distributed time domain signal experiencing rapid variations of its variance, a so called Gaussian transient, can only systematically deviate above unity, while a systematic deviation below unity of the \sk estimator can only be produced by a coherent transient signal, such as artificial RFI contamination.

From this perspective, if the \sk time series illustrated in Figures~\ref{fig:LCP_TDK} and \ref{fig:RCP_TDK} were the result of a high bit sampling \sk analysis, the FRB transient would have been automatically flagged as a coherent terrestrial or astronomical signal. However, given the fact that the background signal demonstrate the same negative \sk-$S_1$ correlation as the FRB transient, the Gaussian nature of the FRB transients cannot be questioned by the result of this analysis. Instead of a rigorous theoretical treatment that would most likely reach to the conclusion that a time domain analysis of 2-bit quantized data would be able to detect but not to discriminate transients signals based on their statistical \sk signature, we present in the following section the results of a set of numerical simulations supporting the same conclusion.

\section{Generalized Spectral Kurtosis analysis of numerical simulated 2-bit quantized transient signals}
\label{sec:sim}
\subsection{Time domain analysis of numerically simulated transients}
To perform the analysis of the numerical simulations presented in this section, we followed a similar procedure as \citet{Nita2016} to generate wide-band constant SNR Gaussian and RFI signals mixed with a normally distributed background, which were turned on only during a compact time segment representing 5\% of entire data stream.

\begin{figure*}[htb]
\begin{minipage}[b]{1.0\linewidth}
  \centering
  \centerline{\includegraphics[width=17.5cm]{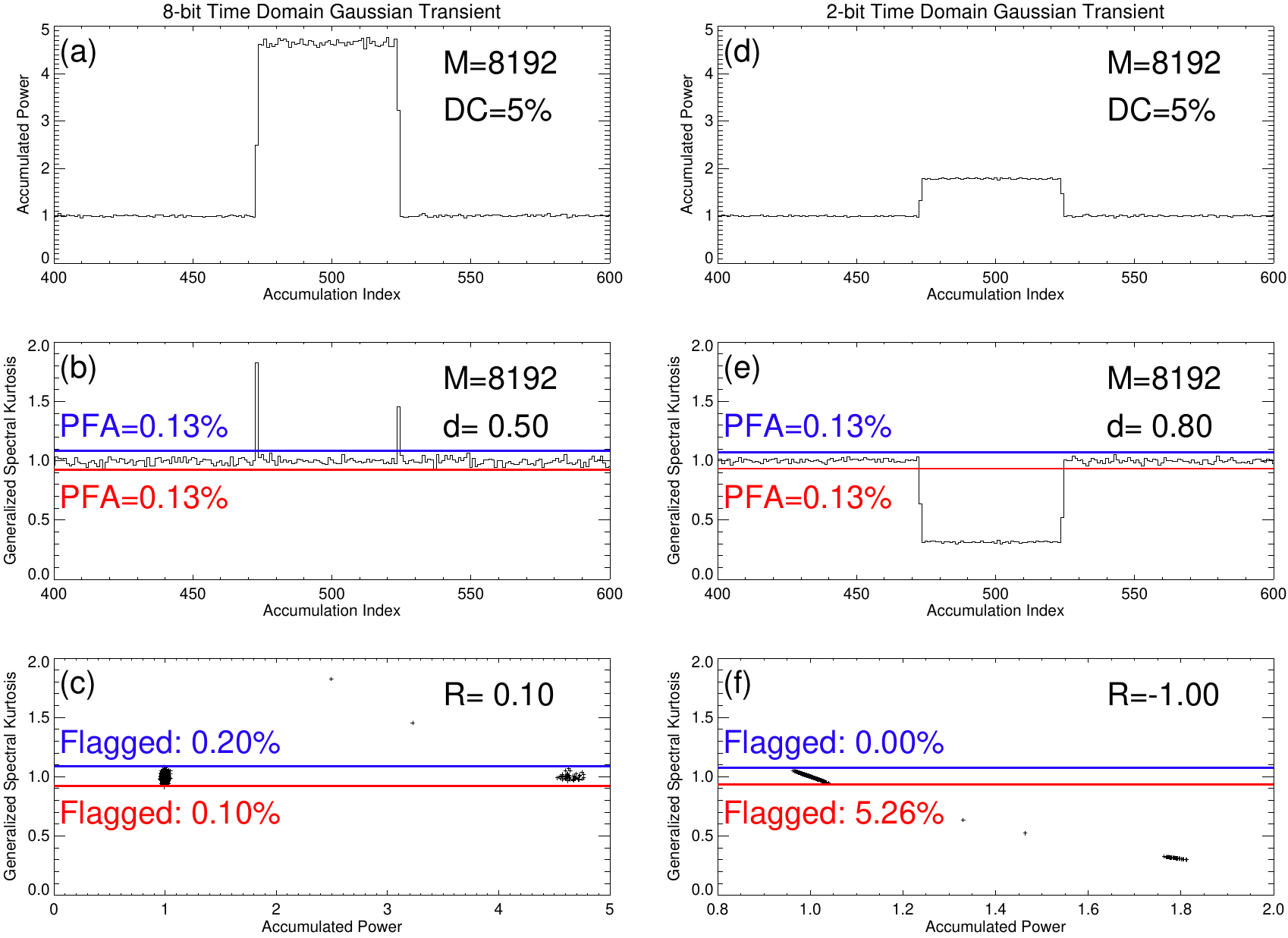}}
 \caption{Time domain analysis of a simulated wide-band Gaussian transient.}
\label{GAUSS_TDK}
\end{minipage}
\end{figure*}

\begin{figure*}[htb]
\begin{minipage}[b]{1.0\linewidth}
  \centering
  \centerline{\includegraphics[width=17.5cm]{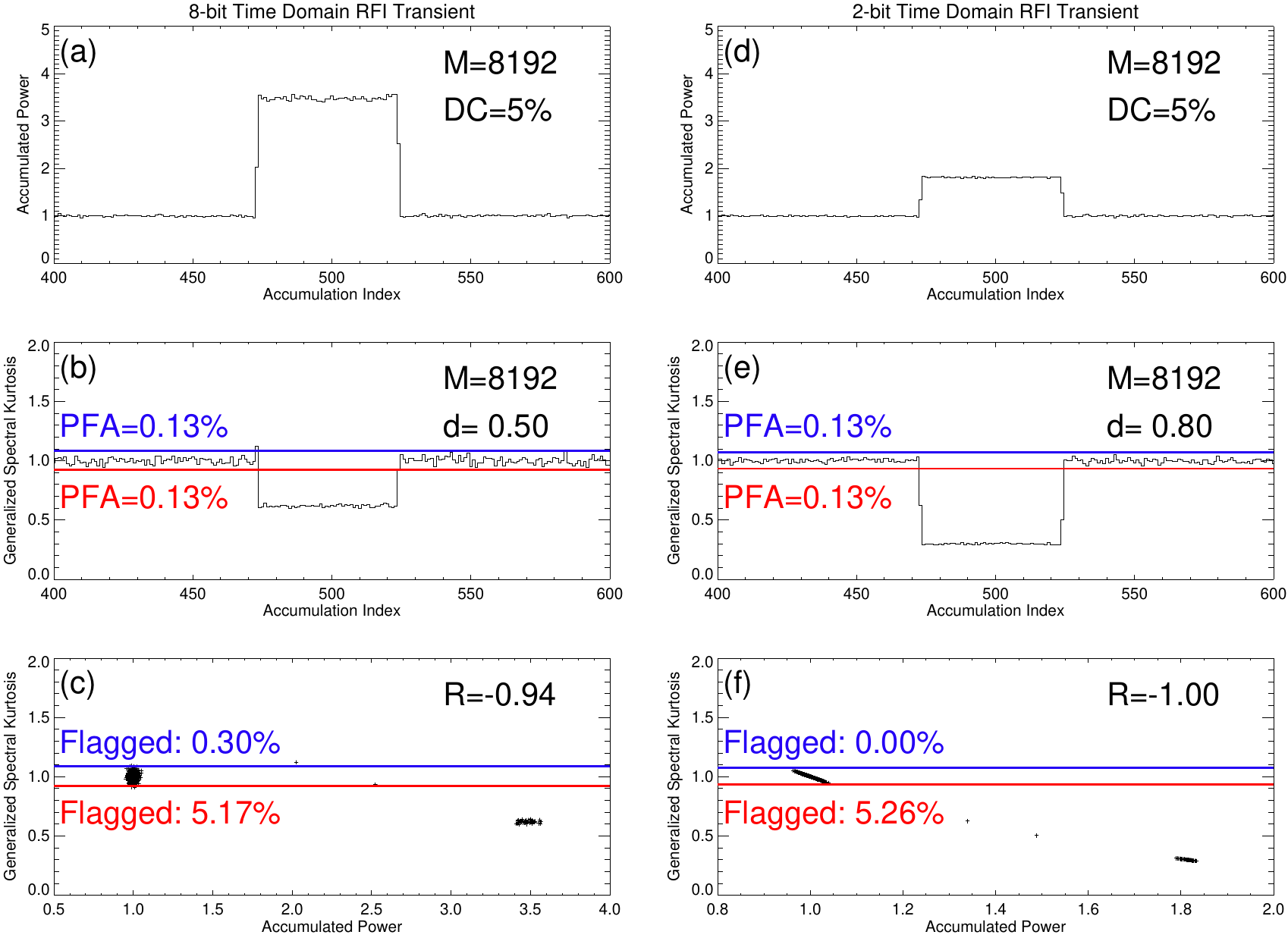}}
 \caption{Time domain analysis of a simulated wide-band RFI transient.}
\label{RFI_TDK}
\end{minipage}
\end{figure*}

For comparison of the \sk performance, the 32-bit numerical data sets were subsequently digitized using 8-bit uniform scaling, as well as a 2-bit quantization scheme optimized for Gaussian signals \citep{qdata}. The 8-bit and 2-bit digitized time series were then integrated over contiguous accumulation blocks of length M=8192 to produce the results illustrated in Figures \ref{GAUSS_TDK} and \ref{RFI_TDK}.

The first column of Figure \ref{GAUSS_TDK} displays a time segment of the 8-bit Accumulated Power ($S_1$), centered on the simulated Gaussian transient, the \sk estimator computed by assuming a ChiSqr distribution of the squared voltage samples ($d=0.5$), and the correlation plot of the \sk-$S_1$ realizations. The \sk time evolution illustrated in panel (b) demonstrates the expected behavior for a high bit sampling estimator, which except for the accumulations capturing the rising and falling edges of the Gaussian transient, evolves within the theoretical 0.13\% PFA non-Gaussianity detection thresholds. The unequal heights of the  two $\esk>1$ spikes marking the edges of the Gaussian transient are due to different duty-cycle of the transient signal relative to the fixed accumulation length \citep{Nita2016}. Also as expected, the panel (c) demonstrates no correlation (R=0.1) between the \sk estimator, which is bounded by the detection thresholds, and the accumulated power $S_1$, which ranges between the background and transient SNR levels.

The 2-bit quantization case, which is illustrated in the second column, demonstrates similarities with the experimental VLBI data analyzed in \S\ref{sec:VLBI_TDK}. Indeed, the\sk estimator, which is normalized to unity by an empirically determined shape factor $d=0.8$, displays a systematic less than unity deviation during the active Gaussian transient, and it is perfectly anti-correlated (R=-1) with the accumulated quantized power corresponding to both background and transient SNR levels. Therefore, the numerical experiment illustrated in Figure \ref{GAUSS_TDK} demonstrates that, in the case of a low-bit sampling data series, a less than unity \sk deviation is the expected behavior for a Gaussian transient, and thus, the results of our time domain \sk analysis presented in the previous section do not exclude the possibility for the FRB 121102 transient to obey a Gaussian statistics.

However, the similar analysis shown in Figure \ref{RFI_TDK} demonstrates that, while the \sk evolution of an RFI transient is clearly distinguishable from the evolution of a Gaussian transient in the 8-bit digitization case, the 2-bit quantization results in a similar $\esk<1$ evolution that makes them indistinguishable by these means. Therefore, the numerical experiment illustrated in Figure \ref{RFI_TDK} demonstrates that a less than unity \sk deviation is also the expected behavior for a non-Gaussian transient, and thus, the results of our time domain \sk analysis presented in the previous section can not unambiguously conclude the true statistical nature of the FRB 121102 transient signal.

\subsection{Spectral domain analysis of numerically simulated transients}
Although the observational data analysis and numerical simulations presented so far clearly demonstrate the limitations of the time domain \sk analysis based on 2-bit quantized voltages, one should recognize that these limitations are mainly the direct consequence of loosing part of the statistical information encoded in the data in the process of producing the 1-bit squared voltages employed in this analysis.

That 2-bit of information may encode enough information to allow statistical discrimination of signals has been clearly demonstrated by \citep{qsk}, who showed that a generalized \sk estimator constructed using the standard power output of the Parkes telescope data pipeline, which consists of on-board accumulations of 2-bit quantized spectral power density estimates, may be used to unambiguously identify the spectral channels contaminated by RFI.

To investigate whether the 2-bit quantized VLBI voltages may actually encode the statistical information needed to unambiguously discriminate different types of transients, we first present in this section the spectral domain \sk analysis of the numerically simulated Gaussian and RFI transients used in the previous section. To proceed with our analysis, and facilitate the direct comparison with the results shown in the previous section, we divided the simulated data in contiguous blocks of length N=128 to produce raw PSD estimates (64 frequency sub-bands) and integrated each M=64 consecutive raw powers ($S_1$ and their squares ($S_2$, to match the same time sampling as in Figures \ref{GAUSS_TDK} and \ref{RFI_TDK}.

\begin{figure*}[htb]
\begin{minipage}[b]{1.0\linewidth}
  \centering
  \centerline{\includegraphics[width=17.5cm]{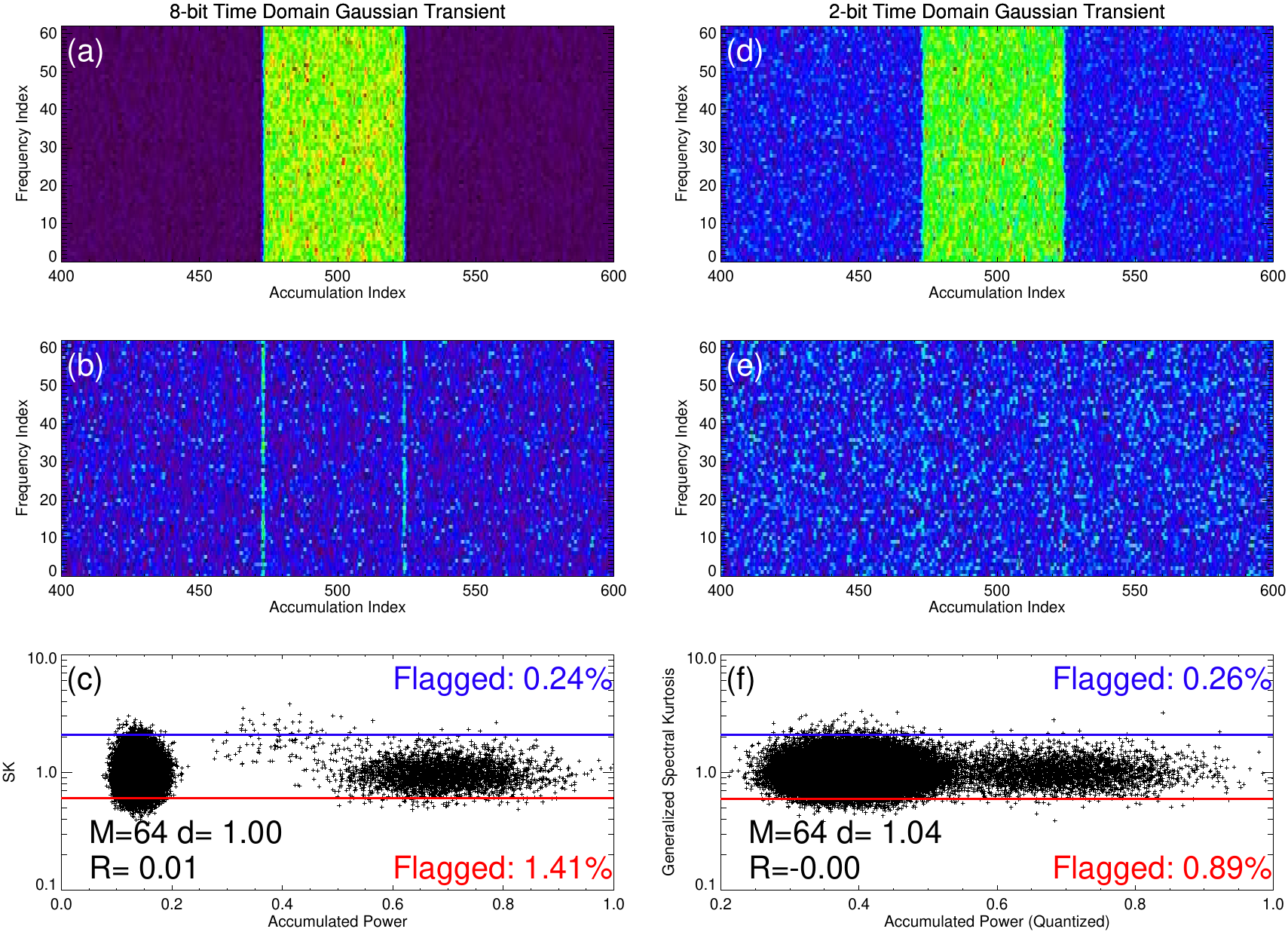}}
 \caption{Spectral domain analysis of the simulated wide-band Gaussian transient presented in Fig.~\ref{RFI_TDK}. The 8-bit and 2-bit quantization results are compared side by side in the left ad right columns, respectively. Top row: Dynamic power spectra (64 frequency channels, M=64 accumulation length). Middle row: \sk dynamic spectra. Bottom row: \sk--$S_1$ correlation plot. The correlation coefficients R shown in each panel demonstrate no correlation. The theoretical 0.13\% PFA detection thresholds above (blue) and below (red) unity expectation resulted in flagging as non-Gaussian time-frequency bins the percentages shown in each panel. The Gamma distribution shape factors used to construct the \sk estimators are also displayed in the bottom panels.}
\label{GAUSS_SK}
\end{minipage}
\end{figure*}

Figures \ref{GAUSS_SK}a and d display the dynamic accumulated power spectra $S_1$ corresponding, respectively, to the 8-bit and 2-bit quantized Gaussian transient data sets. The same as in Figures \ref{GAUSS_TDK}a and d, except for a degraded transient SNR of the 2-bit digitized signal, both spectra clearly reveal the presence of the transient signal. Similarly with the time evolution presented in Figure \ref{GAUSS_TDK}b, the \sk dynamic spectrum shown in panel (b), which was built using the 8-bit quantized data and an assumed shape factor $d=1$, reveals the rising and falling edges of the Gaussian transient, while it fluctuates around the same unity level within and outside its edges. However, panel (d), in which we display the \sk dynamic spectrum built using the 2-bit quantized data and an empirically determined shape factor $d=1$, reveals a remarkable difference in respect to the time evolution shown in Figure~\ref{GAUSS_TDK}d: i.e., the unity normalization of the spectral domain \sk estimator is not sensitive to the presence of the
Gaussian transient signal. Panels (c) and (d) further support this finding, by showing that, unlike in the time domain case, the spectral domain analysis reveals no correlation of the \sk estimator and integrated power $S_1$, for both 8-bit and 2-bit quantizations. Thus, we conclude that, despite the 2-bit quantization of the time domain input, unlike in the time domain, although affected by a degraded sensitivity due to a lower SNR of the active signals, the spectral domain \sk estimator may perform satisfactory well in the presence of Gaussian transients.

\begin{figure*}[htb]
\begin{minipage}[b]{1.0\linewidth}
  \centering
  \centerline{\includegraphics[width=17.5cm]{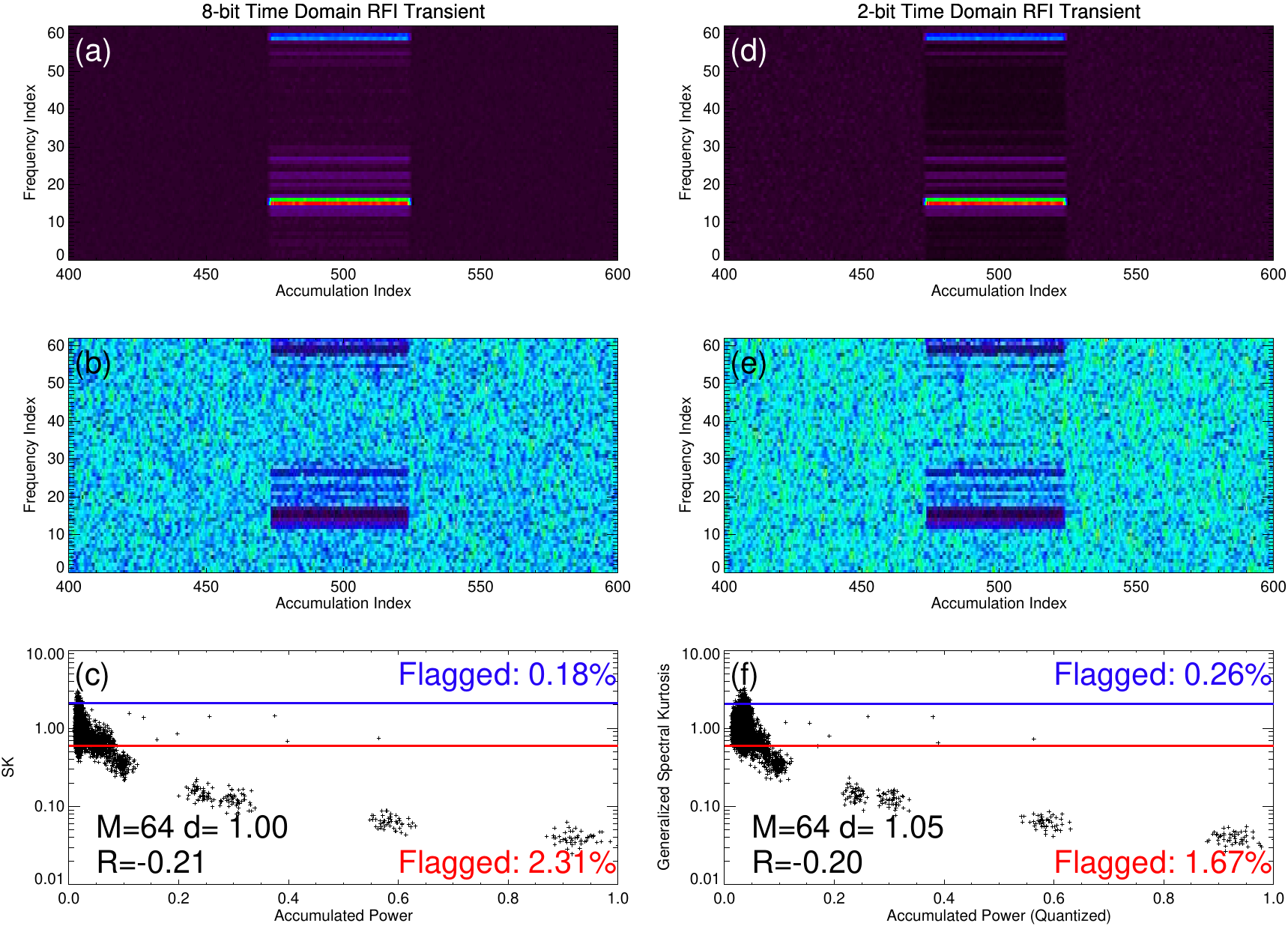}}
 \caption{Spectral domain analysis of the simulated wide-band RFI transient presented in Fig.~\ref{RFI_TDK}. The organization of panels is identical with that in Figure \ref{GAUSS_SK}  }
\label{RFI_SK}
\end{minipage}
\end{figure*}

Similarly, the analysis illustrated in Figure~\ref{RFI_SK} demonstrates no qualitative difference in the the ability of the 8-bit and 2-bit \sk simulators to unambiguously discriminate the statistical signature of the RFI transient that, in both cases, is characterized by by systematic deviations below unity.

To complete this performance analysis of the \sk estimator in the case of quantized transient data, we note that the percentages of flagged data above and below unity are  noticeably larger than the percentages expected from the theoretical 0.13\% PFA detection thresholds, in the case of the Gaussian transient, while, in the case of the simulated RFI transient, they are significantly smaller than the 5\% time domain duty-cycle of the transient. Without performing an in-depth quantitative analysis, in the case of the Gaussian transient we attribute this discrepancy to the quantization process that may affect not only the normalization of the \sk estimator, but also its variance, as it is visually suggested by panels (c) and (f) of Figure~\ref{GAUSS_SK}. As a supporting argument in the favor of this interpretation, we refer the reader to \citet{jai}, where the accuracy of the theoretical detection thresholds was experimentally validated using real astronomical data. In the case of the RFI transient simulation, we attribute the apparent discrepancy between the percentage of flagged RFI data and the actual 5\% time domain duty-cycle, to the fact that, although the wide-band RFI transient was simulated using a constant SNR envelope, as evident in panels (a) and (d) of Figure~\ref{RFI_SK}, the SNR varies across the spectral band, which directly affects the detection performance in the low SNR frequency channels.

Therefore, the analysis of the numerical data presented in this section demonstrates that, despite a low bit sampling time domain data, which may affect its sensitivity, the \sk estimator preserves its ability to discriminate the statistical nature of signals and thus performs in a manner consistent with the general theory developed by \citet{Nita2016} without taking in consideration any quantization effects.

\section{VLBI Spectral Domain Analysis of FRB 121102}
\label{sec:VLBI_SK}
\begin{figure*}[htb]
\begin{minipage}[b]{1.0\linewidth}
  \centering
  \centerline{\includegraphics[width=17.5cm]{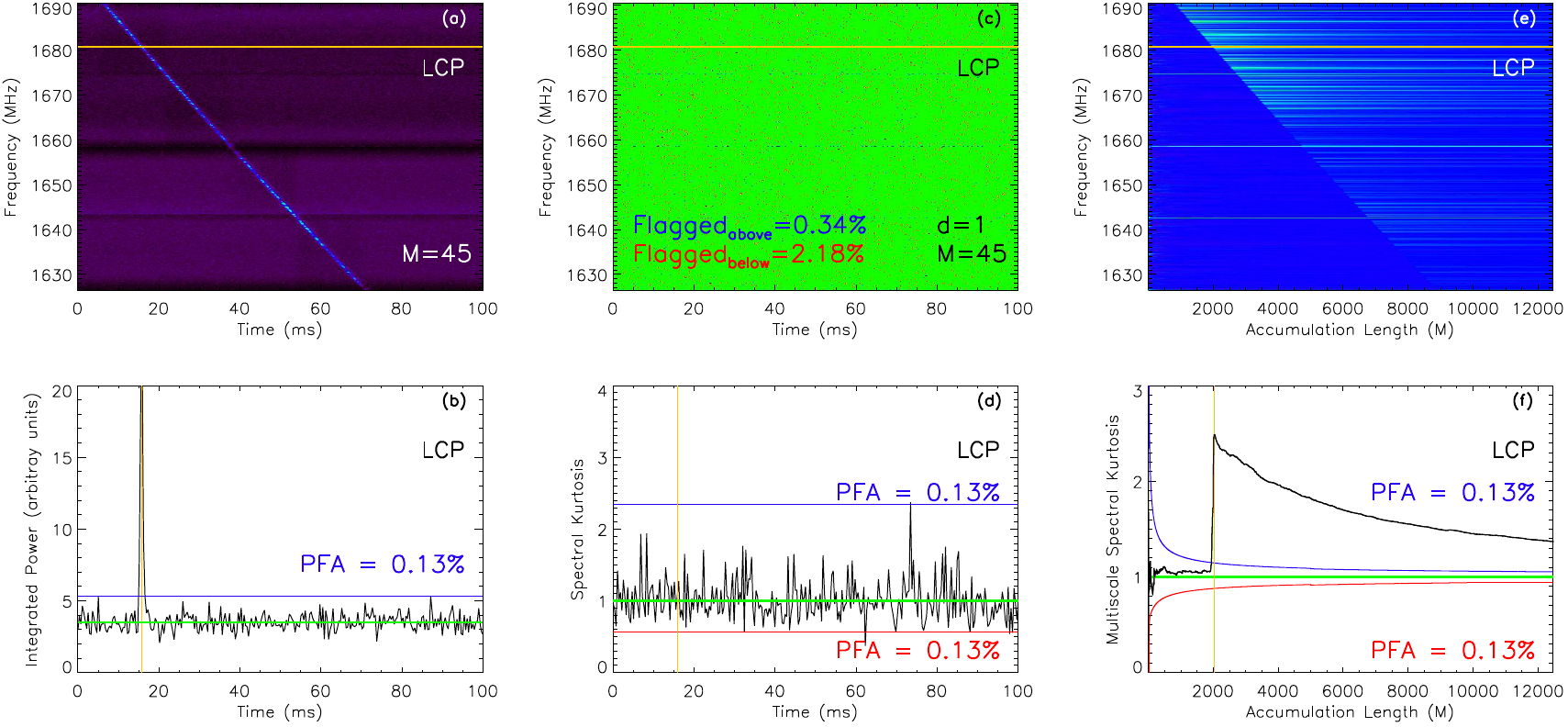}}
 \caption{Spectral analysis of the FRB 121102 VLBI 2-bit sampling LCP voltage data presented in Fig.~\ref{fig:LCP_TDK}: a) Dynamic power spectrum comprising 512 frequency channels (125~kHz width) and accumulations bins of M=45 consecutive raw FFT spectra (0.36~ms time bins). The same time range as in Figs.~\ref{fig:LCP_TDK} and ~\ref{fig:RCP_TDK} has been chosen for display. The dispersed FRB 121102 signal is clearly seen. b) Accumulated power time profile (black curve) corresponding to a selected frequency channel marked by an horizontal yellow line in panel (b). The horizontal blue line indicates the PFA 0.13\% detection threshold above the background level (green horizontal line), assuming the integrated power to be distributed according to gamma distribution of shape factor M=45. The vertical yellow line marks the peak of the detected FRB signal.c) Dynamic spectrum of Spectral Kurtosis (d=1, M=45) flags showing 0.34\% frequency-time bins (blue) located above the upper SK 0.13\% PFA detection threshold and 2.18\% frequency-time bins (red) located bellow the lower SK 0.13\% PFA detection threshold. No systematic flagging associated with the FRB signal seen in panel (a) is observed. d) Spectral Kurtosis time profile (black curve) corresponding to the integrated power light curve shown in panel (b), The upper and lower 0.13\% PFA non-Gaussianity detection thresholds are indicated by the blue and red horizontal lines, respectively. The SK value corresponding to the FRB peak marked by the vertical yellow line in panel (b) is consistent with a time domain Gaussian distribution of the FRB signal. e) Multi-scale Spectral Kurtosis dynamic spectrum obtained by continuous accumulation of the same raw FFT spectra used to build the fixed accumulation length power and SK spectra shown in panels (a) and (c). The presence of the FRB signal is clearly marked by an associated sudden increase of the multi-scale SK values. f) Multi-scale SK time profile (black curve) corresponding to the frequency sub-band marked by horizontal yellow lines in panels (a), (ac) and (e). The blue and red curves indicates the accumulation length dependent 0.13\% PFA detection thresholds above and below the unity SK expectation for a quasi-stationary time domain Gaussian signal. The multi-scale time profile associated with the detection of the FRB signal is consistent with the expected signature of a time-domain Gaussian transient \citep{Nita2016}}
\label{fig:LCP_SK}
\end{minipage}
\end{figure*}

\begin{figure*}[htb]
\begin{minipage}[b]{1.0\linewidth}
  \centering
  \centerline{\includegraphics[width=17.5cm]{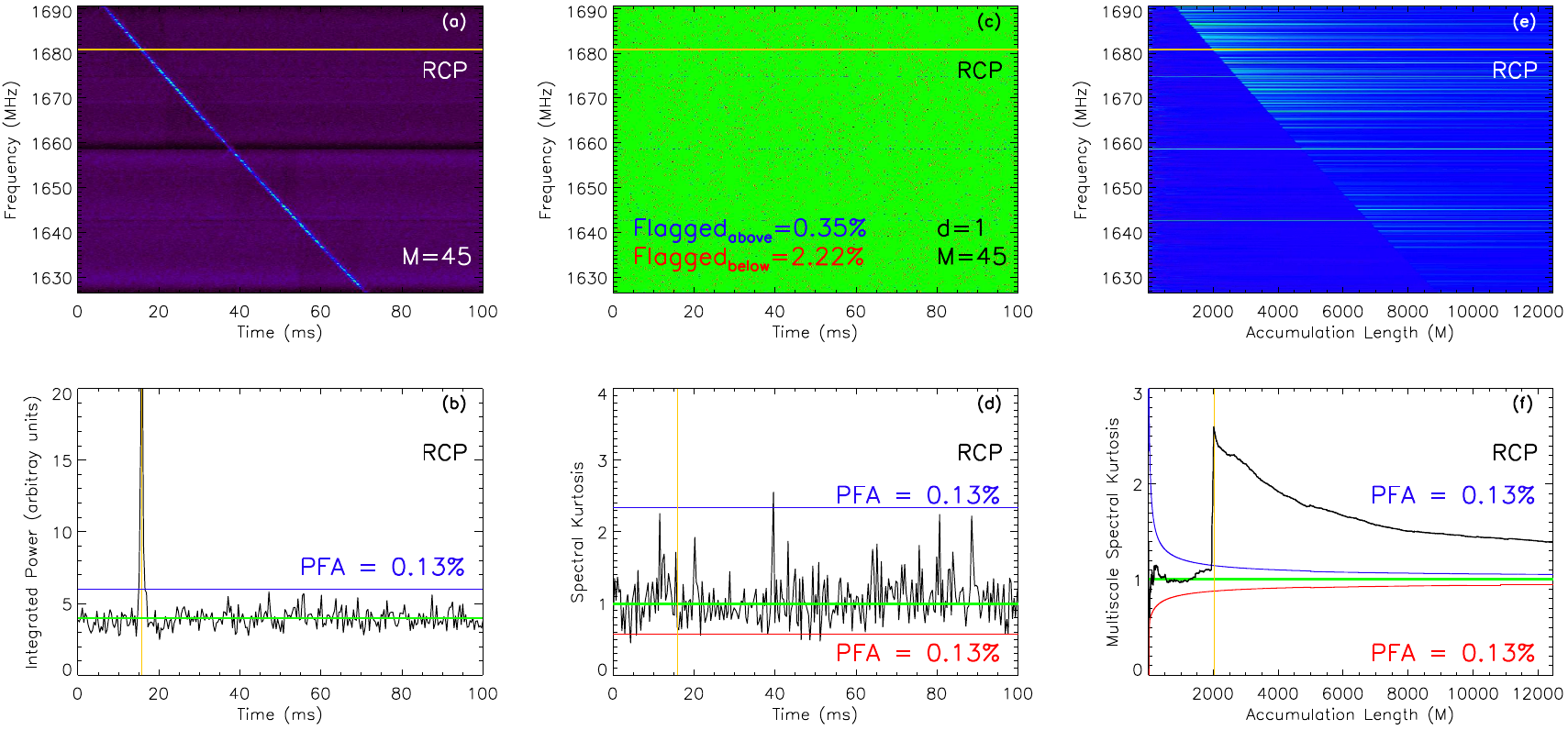}}
 \caption{Spectral analysis of the FRB 121102 VLBI 2-bit sampling RCP voltage data presented in Fig.~\ref{fig:RCP_TDK}.}
\label{fig:RCP_SK}
\end{minipage}
\end{figure*}

To perform the spectral analysis of the FRB 121102 data presented in this section, for each of the 16~MHz sub-bands and both LCP and RCP polarizations, we produced series of raw FFT spectra obtained from 256 consecutive time domain voltage samples. Subsequently, each contiguous block of $M=45$ consecutive raw FFT spectra was accumulated to produce single time bins. This data processing operation resulted in the accumulated power dynamic spectra ($S_1$) shown in  Figures~\ref{fig:LCP_SK}a and \ref{fig:RCP_SK}a, for the LCP and RCP polarizations, respectively, which span the full spectral bandwidth covered by the four observed sub-bands with 125~kHz frequency resolution (512 spectral channels) and 0.36~ms time resolution. To allow a direct comparison with the results presented in \S\ref{sec:VLBI_TDK}, we chose the particular number of channels corresponding to each of the four sub-bands ($N=128$) and the accumulation length ($M=45$) such as to exactly match the time resolution and time span of the time domain accumulations illustrated in Figures~\ref{fig:LCP_TDK} and ~\ref{fig:RCP_TDK}.

In the process of producing the accumulated power spectra $S_1$ shown in Figures~\ref{fig:LCP_SK}a and \ref{fig:RCP_SK}a, we have also produced accumulations of squared raw FFT power spectra, $S_2$, not show here, which in combination to $S_1$, where used to produce the LCP and RCP dynamic \sk spectra displayed in Figures~\ref{fig:LCP_SK}b and \ref{fig:RCP_SK}b, respectively, under the assumption, motivated by the results shown in Figures~\ref{fig:LCP_SK} and \ref{fig:RCP_SK}, that any given raw power sample obtained from a 32-bit FFT applied on 2-bit quantized voltage series obeys an exponential distribution (gamma distribution of shape factor $d=1$) and, consequently, the $S_1$ accumulations obey a gamma distribution of shape factor $M\times d=45$ \citep{sk}.

Remarkably, while the dispersed FRB LCP and RCP signals are clearly seen against the background noise in the accumulated spectra shown in panels (a), the LCP and RCP \sk  dynamic spectra shown in panels \ref{fig:LCP_SK}b and \ref{fig:RCP_SK}b are featureless, which, as illustrated by the numerically simulated Gaussian transient signals shown in Figure~\ref{GAUSS_SK}, may be regarded as the distinct statistical signature of a Gaussian transient with a duration (i.e. $\sim$2~ms) longer than the accumulation time (i.e. 0.36~ms) \citep{Nita2016}.

To quantitatively verify the statistical assumption under which the \sk estimators were built, i.e. $d=1$, we present in panels (b) and (c) of Figures~\ref{fig:LCP_SK} and \ref{fig:RCP_SK}, the single frequency time evolutions of $S_1$ and, respectively \sk, corresponding to a selected frequency channel marked by horizontal yellow lines in the dynamic spectra shown in the top rows of both figures. The $S_1$ LCP and RCP accumulated power profiles shown in panels (b) clearly demonstrate the existence of a practically flat astronomical background, as well as the presence of the bright FRB 121102 LCP and RCP transient signals, whose peaks marked by the yellow vertical lines are located well above the theoretical detection thresholds (blue horizontal lines) relative to the mean background (green horizontal lines), which are expected to be crossed, as the result of statistical fluctuations, only by 0.13\% of a set of random variables obeying a gamma distribution of shape factor $M\times d=45$. Thus, Figures~\ref{fig:LCP_SK}b and \ref{fig:RCP_SK}b, which are representative for the entire data set, indeed demonstrate quantitatively that, unlike the FRB transients, the 32-bit FFT power samples corresponding to the astronomical background are characterized by statistical fluctuations above the mean that are consistent with a gamma distribution having an $M\times d=45$ shape factor.

The same conclusion is also quantitatively supported by the \sk light curves shown in panels (d), which do not cross the lower (red horizontal lines) and upper (blue horizontal lines) 0.13\% PFA thresholds relative to the unity expectation (green horizontal mean) corresponding to an \sk estimator associated with a set of Gaussian time domain samples transformed by the FFT and accumulation operations into a set of random variables that obey a gamma distribution of  $M\times d=45$ \citep{sk}. Moreover, the fact that the \sk values corresponding to the peak times of the LCP and RCP FRB transient signals (marked by yellow vertical lines) do not cross in any direction the 0.13\% PFA non-Gaussianity thresholds, quantitatively supports the conclusion that the the time domain distribution of the FRB transient signal is most likely a Gaussian one.

However, as shown in the panels (c) of both figures, when the thresholding is applied to the entire \sk dynamic spectrum, it is found that $\sim 0.3\%$ of the time-frequency bins cross the upper and $\sim 2.2\%$ cross the lower 0.13\% PFA non-Gaussianity thresholds. The less than unity flagged \sk values indicate contamination by continuous RFI signals or by intermittent RFI signals having a duty-cycle relative to the accumulation length greater than 50\%, which may not be immediately evident in the $S_1$ dynamic spectra shown in panels (a) due to the relatively much larger brightness of the FRB signal. The \sk values crossing the upper non-Gaussianity threshold may be attributed to either $<50\% $ duty-cycle RFI transients or to Gaussian transients of arbitrary duty-cycles \citep{Nita2016}.

Nevertheless, as shown in \citet{Nita2016}, the possibility for the FRB 121102 transient signal to obey a non-Gaussian statistics, which in the case of a such particularly strong SNR could be hidden only by a perfect 50\% duty-cycle, cannot be theoretically dismissed without additional investigations that we perform and illustrate in  Figures~\ref{fig:LCP_SK}(e,f) and \ref{fig:RCP_SK}(e,f).

To perform this investigation, we built the LCP and RCP multi-scale \sk spectra \citep{Nita2016} shown in panels (e) by using the cumulative sums, $S_1(M_i)=\sum_{k=1}^{i} P(k)$ and $S_2(M_i)=\sum_{k=1}^{i} P^2(k)$, and computing the \sk estimators for variable accumulations lengths $M_i=\overline{1,i_{max}}$, where $P$ and $P^2$ represent the frequency--dependent raw FFT estimates, and $i_{max}=45\times 805$ represents the total number raw power estimates used to produce the fixed-length $S_1$ and $S_2$ accumulations (M=45) involved in computing the LCP and LCP multi-scale \sk spectra illustrated in  Figures~\ref{fig:LCP_SK}e and \ref{fig:RCP_SK}e, respectively.

The multi-scale LCP and RCP \sk spectra reveal both the front edge of the dispersed FRB signal, which enters the continuous accumulations at different, frequency-dependent accumulation lengths, and so, different frequency-dependent FRB duty-cycles relative to the progressing accumulation length are realized as the continuous accumulations progress.

Before analyzing the multi-scale \sk profiles shown in panels (f), we note that, beside the FRB edges, the multi-scale \sk spectra also appear constantly brighter than the background at three equidistant frequencies that mark the boundaries between the four sub-bands used to build the composite spectra. This is a theoretically expected behavior, since, in the case of a stationary Gaussian signal, the \sk estimator defined by Equation \ref{gsk} has unity expectation at all but the Nyquist frequencies, where its expectation is 2 \citep{rfi}.

Figures \ref{fig:LCP_SK}f and \ref{fig:RCP_SK}f show the multi-scale \sk profiles at the selected frequency marked by yellow horizontal lines in the top panels. The accumulation length depended 0.13\% PFA thresholds above and below unity are indicated by the blue and, respectively, red curves. The same as in panels (b) and (d), the peak of the FRB LCP and RCP signals are marked by yellow vertical lines. Both LCP and RCP multi-scale \sk profiles have a similar dependence of the increasing accumulation length: they stay within the thresholds until the front edge of the FRB enters the growing accumulation, when \sk has a sudden increase. However, as the accumulation length increase beyond the trailing edge of the transient signals and their relative duty-cycle decreases, the multi-scale \sk profiles asymptotically approach unity, eventually crossing back the upper detection threshold, but never crossing the unity value toward or below the lower detection threshold.

As theoretically shown, and demonstrated by numerical simulations \citet{Nita2016}, such multi-scale \sk profiles may be used to unambiguously determine the Gaussian or non-Gaussian nature of a detected transient, provided that, for a certain range of the accumulation length, the relative duty-cycle of the active signal varies between 50\% and 100\%. However, as it can be inferred from Figures \ref{fig:LCP_SK}a and \ref{fig:RCP_SK}a, only relative duty-cycles much less than 50\% are realized due the choice made for the frequency independent location at which the continuous accumulations are initiated. Therefore, the \sk profiles shown in panels (e) cannot be used to unambiguously infer the true statistical nature of the detected FRB transient.

\begin{figure*}[htb]
\begin{minipage}[b]{1.0\linewidth}
  \centering
  \centerline{\includegraphics[width=15cm]{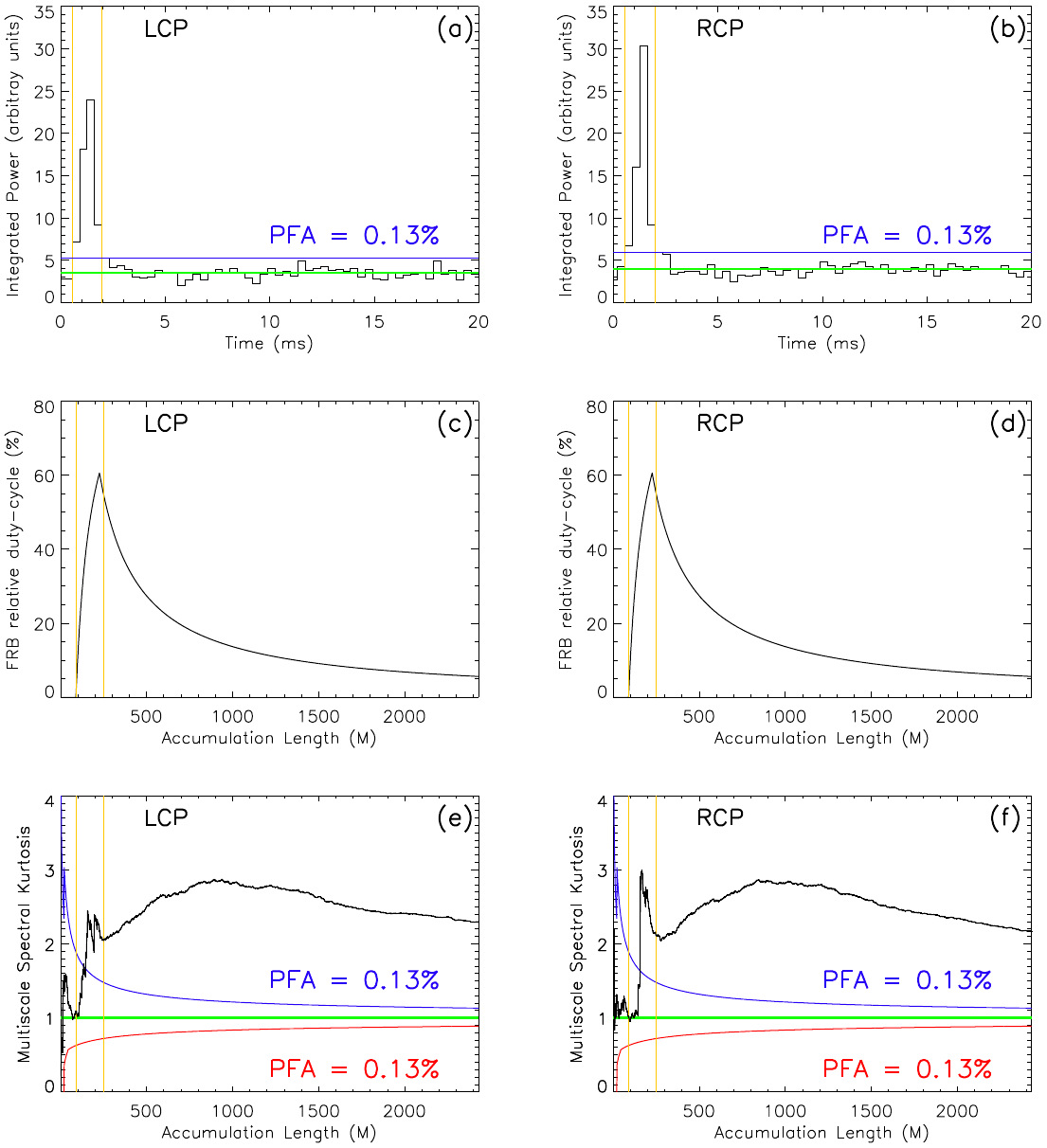}}
 \caption{\sk analysis at the same selected frequency marked by horizontal yellow lines in the upper panels of Figures \ref{fig:LCP_SK} and \ref{fig:RCP_SK}, performed using a multi-scale \sk integration widow starting just before the onset of the FRB signal. Top row: LCP (left panel) and RCP (right panel) accumulated power $S_1$ (M=45 accumulation length). The yellow vertical lines mark the first and the last accumulation bins corresponding to the FRB signals that deviate above the 0.13\% PFA gamma detection threshold (blue horizontal line), relative to the background mean (green horizontal line). Middle row:  Duty-cycle of the active FRB signal relative to the increasing accumulation length. The relative duty cycle is 0\% before the leading FRB edge (first vertical yellow line), rises to $\sim$60\% as the accumulation length increases, and asymptotically drops toward 0\% for accumulations lengths extending beyond the trailing edge (second vertical yellow line) of the active FRB signal. Bottom row: Multi-scale \sk profiles obtained by continuous accumulations of the raw FFT spectra. The leading and trailing edges of the FRB signal are indicated by the two vertical yelow lines and the upper and lower 0.13\% non-Gaussianity detection thresholds are indicated by the blue and red curves, respectively.}
\label{fig:SK_udd}
\end{minipage}
\end{figure*}

However, once the transient signal is detected, relative duty-cycles larger than 50\% can be realized by employing a sliding accumulation widow that, for a given frequency, is conveniently positioned to start the continuous accumulations close enough to the trailing edge of the transient signal. The result of this approach is illustrated, for both polarizations, in Figure \ref{fig:SK_udd} for the same selected frequency as in Figures \ref{fig:LCP_SK} and \ref{fig:RCP_SK}.

As indicated by the two vertical yellow lines in panels (a) and (b), the short duration active FRB signal ($\sim$2~ms) is detected in four consecutive accumulation bins ($\sim$180 raw FFT spectra), which are preceded by one accumulation bin (45 raw FFT spectra) containing only the background signal. Therefore, as shown in the middle row panels of Figure~\ref{fig:SK_udd}, the relative duty-cycle of the transient signal relative to the increasing accumulation length, defined as the ratio of raw FFT spectra containing the active signal to the total number of accumulated raw FFT spectra,  is identically 0\% before the leading FRB edge, rises to 60\% as the accumulation length increases, and asymptotically drops toward 0\% for accumulations lengths extending beyond the trailing edge of the active signal. This duty-cycle variation results in the LCP and RCP multi-scale \sk profiles shown in panels (e) and (f), respectively. With this specifically tailored integration setup, both \sk profiles cross the upper detection threshold to reach a first peak within the boundaries of the transient FRB signal, decrease to a local minimum larger than unity that corresponds to the trailing edge of the transient, gradually increase to reach a second local maximum, and eventually decrease toward unity as the transient relative duty-cycle asymptotically decrease toward 0\%.

As demonstrated by \citet{Nita2016}, this \sk dependence on the accumulation length (and the corresponding relative duty-cycle) confirms the Gaussian time domain statistics of the FRB 121102 signal, as opposed to an hypothetical coherent (RFI) transient that, distinctively, would have crossed the unity level, for duty-cycles ranging from 50 to 60\%, to reach a less than unity minimum value before rising again to reach the second, larger than unity, local maximum.

Therefore, the multi-scale \sk analysis illustrated in Figure \ref{fig:SK_udd} unambiguously demonstrates that the FRB 121102 signal obeys a Gaussian statistics.

\section{Conclusions}
\label{sec:conclusions}
We have presented a statistical analysis based on the generalized \sk estimator of a 2-bit quantized time domain series containing the FRB 121102 transient signal, which was recorded in both LCP and RCP polarizations by the EVN instruments.

We demonstrated that the low bit sampling affects the time-domain statistical properties of the recorded voltages, however, the generalized \sk estimator may be still employed for the purpose of detecting the presence of natural or artificial transient signals mixed in the time domain sub-bands, although they cannot be distinguished from each other.

Nevertheless, we have also demonstrated that, after a 32-bit FFT applied to the 2-bit quantized voltages, the true statistical nature of the signal is recovered, which allows unambiguous statistical discrimination between natural and artificial transients. This finding indicates that, in contrast with any power-based, empirical, thresholding algorithm that, by ignoring the higher order statistical information encoded in the signal, would unselectively interpret any detected transient signal as RFI contamination, the spectral domain \sk estimator can be safely and efficiently employed for the purpose of automatic detection and selective RFI removal in such 2-bit quantized data streams

Finally, we performed a multi-scale \sk analysis of the recorded signals that unambiguously determined the Gaussian nature of the FRB 121102 transient, this being, on best of our knowledge, the first ever study that reaches such conclusion based on \sk statistical analysis. However, our conclusion pertains only to the particular FRB 121102 signal we have analyzed, thus, the possibility for other FRB signals to obey a non-Gaussian statistics cannot be excluded without a case by case analysis.

\section*{Acknowledgements} This work was partly supported from the NSF grant AST-1615807 to the New Jersey Institute of Technology.


\end{document}